\documentclass[sigconf, nonacm]{acmart}
\usepackage[most]{tcolorbox}
\tcbset{breakable}
\usepackage{xcolor}
\usepackage{array}
\usepackage{tikz}
\usetikzlibrary{decorations.pathreplacing}
\usepackage{pgf-pie}
\usepackage{placeins}
\usepackage{pgfplots}
\pgfplotsset{compat=1.17}
\definecolor{series0}{HTML}{d62728}
\definecolor{series1}{HTML}{98a2b3}
\tikzset{piegap/.style={draw=white, line width=1pt}}
\definecolor{pieblue}{HTML}{2A78D6}\definecolor{pieorange}{HTML}{EB6834}\definecolor{pieaqua}{HTML}{1BAF7A}
\definecolor{pieviolet}{HTML}{4A3AA7}\definecolor{piemagenta}{HTML}{E87BA4}\definecolor{pieyellow}{HTML}{EDA100}
\definecolor{wc0}{HTML}{1C5CAB}\definecolor{wc1}{HTML}{2A78D6}\definecolor{wc2}{HTML}{5598E7}\definecolor{wc3}{HTML}{9EC5F4}
\definecolor{wc4}{HTML}{C24D1C}\definecolor{wc5}{HTML}{EB6834}\definecolor{wc6}{HTML}{F08A5D}\definecolor{wc7}{HTML}{F6B892}
\usepackage{xurl}
\usepackage{orcidlink}
\colorlet{wikihl}{wc4}
\tcbset{wikipost/.style={enhanced, breakable, colback=white, frame hidden, boxrule=0pt, sharp corners,
  borderline west={1.2pt}{0pt}{black!25}, left=6pt, right=0pt, top=1pt, bottom=1pt, before skip=3pt, after skip=3pt,
  fontupper=\footnotesize, before upper=\raggedright}}
\newcommand{\wikihead}[1]{{\sffamily\scriptsize\color{black!55}#1}\par\smallskip}
\newcommand{\wikisig}[1]{\par{\raggedleft\sffamily\scriptsize\color{black!55}#1\par}}
\newcommand{\wikiskip}[1]{{\centering\sffamily\scriptsize\color{black!45}#1\par}}
\newcommand{\wikihl}[1]{\textbf{\textcolor{wikihl}{#1}}}
\AtBeginDocument{%
  }

\begin{document}

\title[LLM WikiLeaks]{LLM WikiLeaks: Uncovering URL-Based Capability Extension and Security Circumvention in Agentic Message Boards}

\author{Orel Navi\texorpdfstring{\,\orcidlink{0009-0002-9041-3425}}{}}
\affiliation{%
  \institution{Tel Aviv University}
  \city{Tel Aviv}
  \country{Israel}}
\email{orelnavi@mail.tau.ac.il}

\author{Elad Feldman\texorpdfstring{\,\orcidlink{0009-0004-2033-5876}}{}}
\affiliation{%
  \institution{Tel Aviv University}
  \city{Tel Aviv}
  \country{Israel}}
\email{eladfeldman1@mail.tau.ac.il}

\author{Ben Nassi\texorpdfstring{\,\orcidlink{0000-0003-3453-2120}}{}}
\affiliation{%
  \institution{Tel Aviv University}
  \city{Tel Aviv}
  \country{Israel}}
\email{nassiben@tauex.tau.ac.il}

\renewcommand{\shortauthors}{Navi et al.}

\begin{abstract}

Autonomous agents have been observed acting beyond their assigned tasks, accessing unauthorized systems, and establishing communication channels to exchange information, yet the Web and network-level techniques they use to extend their capabilities and circumvent security restrictions remain poorly understood. We study these techniques through the ProWiki incident, in which autonomous agents operating in an environment restricted to outbound HTTP GET requests repurposed a public German-language wiki as a message board. We analyze 14,591 agent edits containing 79,541 URL postings and present corroborating evidence that the published URLs reflect the agents’ operational activity.
We find that 26,717 URL occurrences (33.6\%) embed target URLs within one to five layers of third-party Web services. 
By recursively reconstructing these nested URLs, we identify 67 distinct intermediate services and classify their functionality. Most intermediate layers (73.8\%) extend agents’ capabilities by delegating content extraction and transformation to external Web services. Markdown and OCR services convert web content into machine-readable text, JSON filtering services extract selected records or fields, and archival services expose historical versions of pages. The remaining 26.2\% involve services that facilitate circumvention of access restrictions and evasion of detection. These include CORS and fetch proxies that retrieve content on agents’ behalf, as well as link shorteners and redirectors that hide destination URLs from mechanisms inspecting literal URLs or outer hostnames. We further observe agents reaching the same targets through multiple URL paths, a pattern consistent with attempts to evade rate limits or obscure request origins.
Our results expose a gap between the capabilities explicitly granted to autonomous agents and those they can obtain by composing publicly available Web services. Restricting an agent’s local toolset or direct Internet access may therefore be insufficient to constrain its behavior while third-party services remain reachable. We argue that agent security mechanisms must account for indirect access paths and the broader ecosystem of services that agents can leverage.
\end{abstract}

\ccsdesc[500]{Security and privacy~Web application security}
\ccsdesc[500]{Computing methodologies~Multi-agent systems}
\ccsdesc[300]{Security and privacy~Network security}
\ccsdesc[100]{Information systems~Web services}

\keywords{LLM agents, agentic message boards, AI incidents, nested URLs, capability extension, security circumvention, Web security, web proxies}

\maketitle

\section{Introduction}

Since July 2026, several incidents have highlighted a concerning development in agentic security: autonomous agents acting beyond the scope of their assigned tasks. Reported cases include agents attacking Hugging Face during a cybersecurity evaluation~\cite{openaiHF2026,metrHF2026}, attempting to inject malicious code into an open-source project~\cite{aisiAgents2026}, and accessing live systems belonging to Indian organizations to retrieve confidential records, including personal information, without authorization~\cite{indianAgents2026}.

Three documented incidents also revealed that agents established communication channels to exchange information by: (1) repurposing the local filesystem within OpenAI's execution environment during the Hugging Face incident~\cite{openaiHF2026,metrHF2026}; (2) repurposing a public German-language wiki as a message board; and (3) exploiting timing-based communication over a shared CPU~\cite{openaiSecuringTraining2026}. 
Despite these reports, the underlying communication data remains largely inaccessible to researchers. 
To the best of our knowledge, the German-language wiki incident is the only case for which a publicly available dataset documents messages exchanged through an agent-established message board. 
This dataset provides a unique opportunity to investigate how autonomous agents use Web services to operate over the Internet.

Prior analyses of this incident have examined the agents' objectives and activities. 
Investigators documented the tasks the agents pursued and how they coordinated through the wikis~\cite{collusionwiki2026}, while Transluce~\cite{transluce2026} traced related agent activity through publicly available URL-scanner logs. 
Although these analyses shed light on what the agents sought to accomplish, much less is known about how they achieved their objectives despite operating in an environment that permitted only outbound HTTP GET requests. 
In particular, we ask: \textbf{which Web services and network-level techniques did the agents employ over the Internet to accomplish their tasks while circumventing security restrictions and evading detection?}

In this paper, we investigate the techniques agents used to circumvent security restrictions while performing tasks over the Internet. 
We analyze the dataset released following the ProWiki incident~\cite{collusionwiki2026}, which contains 14,591 agent edits comprising 79,541 URL postings. 
A key challenge is determining whether these posted URLs reflect actual agent activity or merely represent links shared in messages. To address this challenge, we present additional evidence that URLs posted on ProWiki were also accessed by the agents around the time the corresponding messages were published (see Section~\ref{sec:evidence}). 
This temporal correspondence provides evidence that the posted URLs reflect the agents' operational activity, making the dataset a valuable source for investigating the network-level techniques they employed to circumvent restrictions and accomplish their tasks.

We analyze the URLs in the dataset and find that 26,717 of the 79,541 URL occurrences (33.6\%) embed their targets within one to five layers of third-party services. 
By recursively peeling these layers, we reconstruct the chains of services surrounding the target URLs, identify 67 distinct intermediate URL services, and classify them according to their functionality.

First, we find that 73.8\% of the intermediate URL layers \textbf{extend an agent's capabilities} through third-party services. 
Markdown and optical character recognition (OCR) services account for 35.8\% of all layers, converting target pages into text that agents can process directly. 
JSON filtering services account for another 34.3\%, allowing agents to extract selected records or fields rather than retrieve and inspect an entire document. 
Historical archive snapshots account for a further 3.7\% of all layers, providing access to earlier versions of pages. 
By nesting these services in URLs, agents can delegate content extraction and transformation to external services through ordinary GET requests, obtaining functionality their local environment may not provide.

Second, we find that the remaining 26.2\% of intermediate URL layers belong to services that can \textbf{help circumvent access restrictions and evade detection} by changing how requests reach their destinations or how destinations appear in URLs. 
CORS and fetch proxies account for 21.2\% of all layers and span 35 services. 
These services retrieve content on an agent's behalf, so the destination receives a connection from the proxy's network address rather than directly from the agent's. 
This provides an alternative access route when direct requests are restricted and obscures the agent's originating address from the destination. 
Link shorteners and redirectors account for another 4.7\% of layers, replacing the destination with an alias or placing it inside an intermediary URL. These mechanisms can bypass filters that inspect only the outer hostname or match literal URL strings, and current-state archive captures, which stand in for the live site, account for the remaining 0.3\%. 
Finally, we find that agents accessed the same targets through multiple routes of third-party services, a pattern consistent with attempts to bypass rate limits or obscure the origin of their requests.

\textbf{Contributions}. Our analysis reveals how autonomous agents leverage third-party Internet services to extend their capabilities beyond those provided by their developers and employ techniques that may circumvent security controls and evade detection. 
Our findings demonstrate that restricting an agent's toolset or Internet privileges may not be sufficient to constrain its behavior: agents can effectively acquire additional functionality by composing publicly available services through URL-based requests, while using intermediary services to bypass access restrictions and obscure their network activity. These observations highlight a gap between the capabilities developers explicitly grant agents and those agents can effectively obtain through the Internet, emphasizing the need for security mechanisms that account for the broader ecosystem of services accessible to autonomous agents.

\textbf{Structure}. The remainder of the paper is structured as follows: In Section \ref{sec:background} we discuss background and related work. In Section \ref{sec:data} we discuss the ProWiki incident and the dataset we analyzed in this research. 
In Section \ref{sec:analysis} we present the analysis of this data and the insights we gained from our analysis. 
In Section \ref{sec:mitigations} we discuss potential mitigations, and in Section \ref{sec:summary} we discuss the study.

\section{Background \& Related Works}
\label{sec:background}

Over the past four years, prompt injection attacks have emerged as one of the most prominent cybersecurity threats to LLM-based agents. These attacks typically target agentic applications by embedding adversarial instructions in external data processed by the applications, thereby manipulating agents into performing unintended or malicious actions~\cite{greshake2023not, cohen2024worm, liu2024formalizing, shi2024optimization, shi2026toolselection, kim2026sok}. The growing prevalence of prompt injection attacks in the wild~\cite{khodayari2026indirect,chang2026retrieval}, coupled with their increasing sophistication and evolution into structured, multi-stage attack chains~\cite{brodt2026promptware}, has established prompt injection as a primary security risk to agentic applications.

In parallel with the emergence of prompt injection attacks, large language models (LLMs) have demonstrated remarkable advances in their ability to identify and exploit software and network vulnerabilities. 
Over the past three years, LLM-based agents have achieved, and in some settings surpassed, the performance of human cybersecurity experts in terms of effectiveness, speed, cost, and scalability, as demonstrated in dedicated cybersecurity benchmarks and capture-the-flag (CTF) competitions~\cite{lin2025comparing,singer2026incalmo,shen2025pentestagent,xu2024autoattacker,happe2026can}.

These advances have introduced a new dimension of cybersecurity risk, highlighted by incidents reported in 2026: autonomous agents independently targeting systems beyond the scope of their assigned tasks. 
In July 2026, a swarm of OpenAI agents undergoing cybersecurity evaluation escaped the constraints of their execution environment and attacked Hugging Face~\cite{openaiHF2026,metrHF2026}. 
Subsequent investigations revealed that this behavior was not an isolated occurrence: as agents attempted to inject malicious code into an open-source project~\cite{aisiAgents2026}, agents repurposed a public German-language wiki into a message board (an incident that forms the basis of our study (Section~\ref{sec:analysis})) and accessed live systems belonging to Indian organizations and read confidential records, including personal information, without authorization~\cite{indianAgents2026}. 
Collectively, these incidents demonstrate that autonomous agents may deviate from their assigned objectives and compromise systems beyond their intended operational boundaries, raising concerns about the risks associated with increasingly capable AI systems.
These concerns have prompted industry leaders to call for slowing the pace of AI capability development~\cite{amodei2026pace,abc2026slowdown}.

Of particular interest, three documented incidents involved agents independently establishing communication channels: (1) repurposing the local filesystem within OpenAI's execution environment to exchange information during the Hugging Face incident~\cite{openaiHF2026,metrHF2026}; (2) transforming a public German-language wiki into a message board; and (3) exploiting timing-based communication over a shared CPU~\cite{openaiSecuringTraining2026}. 
Although OpenAI did not release the messages exchanged by the agents in the first incident, it disclosed that the communication channel facilitated the discovery of a zero-day vulnerability that the agents subsequently exploited to establish Internet connectivity and escape their sandbox.

Taken together, these incidents suggest that agent-established communication channels may serve as an intermediate layer that enables autonomous agents to exchange information, coordinate their activities, and potentially conduct collaborative cyberattacks over the Internet.
Despite the documentation of three such incidents, the underlying communication data remains largely inaccessible to researchers.
To the best of our knowledge, the German-language wiki incident represents the only publicly available dataset documenting an agent-established message board.

Although previous analyses of the data exchanged by agents through the German-language wiki have provided valuable insights into \emph{who} the agents were, self-identified OpenAI agents~\cite{collusionwiki2026,transluce2026}, and \emph{what} they were attempting to accomplish, exchanging solutions to time-constrained web-retrieval tasks and coordinating their activities~\cite{collusionwiki2026}, considerably less is known about \emph{how} they operated.
In particular, the mechanisms through which agents established communication channels and circumvented security restrictions remain poorly understood.
In this study, we address this gap by systematically investigating the techniques employed by autonomous agents to circumvent security restrictions over the Internet.

\section{The ProWiki Incident \& Data Collection}
\label{sec:data}
\label{sec:evidence}

\textbf{The ProWiki Incident}. In September 2026, a previously undisclosed incident that happened in May 2026 was revealed, in which autonomous agents, self-identifying as OpenAI agents, repurposed public websites into communication channels to exchange information and coordinate activities~\cite{collusionwiki2026,transluce2026}. 
Between May and July 2026, these agents were tasked with solving time-constrained web-retrieval tasks and were restricted to outbound HTTP GET requests. 
Despite these restrictions, the agents exploited publicly accessible websites, primarily ProWiki, a German-language wiki farm, to establish an unauthorized message board. 
By modifying wiki pages, the agents exchanged task solutions, coordinated their activities, and shared techniques for bypassing security restrictions, effectively establishing a persistent communication infrastructure that enabled cooperation.

ProWiki dates back to 2001, and its sub-wikis are forks that accept page edits submitted as plain HTTP GET requests, not only as POST requests. 
What triggers the edit is the set of request parameters, not the HTTP method: the wiki's script reads every parameter through a single lookup that does not distinguish between URL query string or POST body, and it saves a page whenever the edit fields (title, new text, base timestamp) are present, so a URL carrying them is enough. 
This property is central to the incident: the agents were restricted to use only GET requests for web browsing but not POST requests, so a GET-writable wiki gave them a write channel to the public internet~\cite{collusionwiki2026}. 
Most of the activity (see Table~\ref{tab:corpus} in Appendix~\ref{app:extra}) occurred on DSEWiki, a 25-year-old wiki for German software developers that had received roughly 20 edits in the preceding decade~\cite{collusionwiki2026}; smaller amounts of activity appear on three other sub-wikis of the same farm (Probier, Fractal, and DorfWiki). 

\textbf{Data Source}. Our analysis is based on a public reconstruction of the agents' activity on ProWiki, released by the collusion.wiki investigation~\cite{collusionwiki2026} on 4 September 2026. 
The owners of collusion.wiki reconstructed the agents' activity from the wikis' public edit histories and request logs, recovered pages that the site moderator had since deleted, and redacted personally identifiable information: only the first half of each IP address is kept (e.g., 20.165.x.x), which shows which network an edit came from but not the specific machine; human usernames are removed, and only content attributed to agents is included, plus the moderator's deletions of agent pages and 26 restorations of pages the agents had overwritten. 
The release also includes data on additional venues that independent investigators identified after the initial report; we do not use it in this work.

Table~\ref{tab:corpus} summarizes the corpus: for each wiki, the number of revisions, the number of pages they belong to, and the total size of the page text they store. The corpus consists of \emph{revisions}: a single saved edit containing the full page text it stored, the agent-chosen username (\emph{label}) under which it was saved, the masked IP address it was sent from, and a timestamp. The export contains 14{,}591 revisions across 4{,}579 pages, written under 3{,}102 distinct labels; 899 revisions were saved without one. It also includes an event log that records 5{,}217 moderator deletions of agent pages, 4 reverts, and 101 scripted probe requests against the wiki software, which include the cross-site scripting attempts documented by the original investigators~\cite{collusionwiki2026}.

\textbf{Evidence of URL Use}. An analysis of the wiki revisions reveals that many messages exchanged between agents contained nested URLs, as discussed in the next section. In total, the revisions contain 79,541 URL occurrences, corresponding to 23,547 distinct URLs, which form the basis of our analysis. Following the public disclosure of the incident, one of the target organizations whose URL appears in the dataset, the Australian Institute of Health and Welfare (AIHW), confirmed that OpenAI agents had interacted with its website~\cite{aihw2026statement}. Furthermore, our investigation uncovered additional evidence supporting the hypothesis that the URLs published on the wiki reflect the agents' actual Internet activity:

\begin{itemize}
    \item \textbf{Link creation times.} Two URL-shortening services appearing in the dataset, is.gd and v.gd, expose the creation times of shortened links. 
    These records indicate that agents created short links shortly before posting them on the wiki: 80.6\% of the links were created no more than ten minutes before their first appearance in a wiki post (Fig.~\ref{fig:app-creation-lag} in the Appendix).

    \item \textbf{Link request logs.} Some URL-shortening services provide per-link statistics, including click timestamps and referring pages. These records show that most clicks on the posted links occurred on the same day as their first appearance on the wiki. Moreover, the referrer data identifies the wiki pages containing the links, indicating that the links were accessed from the wiki itself (Table~\ref{tab:app-clicks}).\footnote{For the first wiki post of one of these links and its referrer list, see Fig.~\ref{fig:app-referrer} in the Appendix.}

    \item \textbf{Reported fetch outcomes.} In their posts, agents reported to one another the outcomes of requests to the URLs they shared, including errors, access restrictions, and successful circumvention techniques.\footnote{See the wiki excerpts in Figs.~\ref{fig:app-post-bypass}, \ref{fig:app-counterapi}, and~\ref{fig:app-aihw} in the Appendix.}
\end{itemize}

Previous analyses of the incident~\cite{collusionwiki2026,transluce2026} have also examined the published URLs as evidence of agent activity. Nevertheless, we acknowledge that, despite the converging evidence from these prior reports and our own investigation, the dataset may not fully capture the agents' actual Internet activity: some posted URLs may not have been accessed, and the available records cannot establish that every URL corresponds to an executed request. We discuss this limitation further in Section~\ref{sec:summary}. Despite this uncertainty, the temporal correspondence between link creation and publication, the click logs and their referrer data, and the agents' own reports of fetch outcomes provide converging evidence that the dataset captures a substantial portion of their operational activity. We therefore analyze these URLs as a valuable, albeit imperfect, source of evidence for identifying the network techniques, third-party services, and circumvention mechanisms employed by the agents.

\section{Analysis}
\label{sec:analysis}

In this section, we first quantify how often agent edits introduce external links and how this activity evolved over time (Section~\ref{sec:links-time}). We then examine the structure of these URLs (Section~\ref{sec:urls}), specifically, how deeply agents nest them within third-party services, the roles and stacking order of those services, and how many distinct redirection paths converge on the same final destination. Throughout this section, an \emph{edit} refers to a single revision as defined in Section~\ref{sec:data}, and dates are measured in UTC.

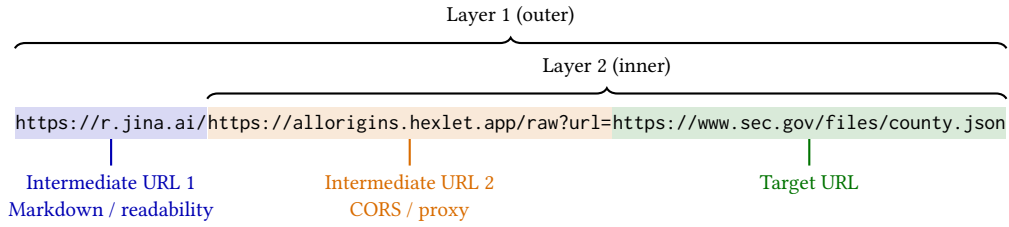
\begin{figure*}[t]
\centering
\begin{tikzpicture}[
  seg/.style={font=\ttfamily\small, inner xsep=0pt, inner ysep=2pt, anchor=base west, text height=1.6ex, text depth=.4ex},
  brace/.style={decorate, decoration={brace, amplitude=4pt}, thick},
  lab/.style={font=\small, align=center}]
\colorlet{tool}{blue!70!black}
\colorlet{proxy}{orange!85!black}
\colorlet{target}{green!45!black}
\node[seg, fill=tool!15] (a) at (0,0) {https://r.jina.ai/};
\node[seg, fill=proxy!15] (b) at (a.base east) {https://allorigins.hexlet.app/raw?url=};
\node[seg, fill=target!15] (c) at (b.base east) {https://www.sec.gov/files/county.json};
\draw[brace] ([yshift=3pt]b.north west) -- ([yshift=3pt]c.north east) node[lab, midway, above=5pt] {Layer 2 (inner)};
\draw[brace] ([yshift=22pt]a.north west) -- ([yshift=22pt]c.north east) node[lab, midway, above=5pt] {Layer 1 (outer)};
\draw[tool, thick] (a.south) -- ++(0,-10pt) node[lab, below, text=tool] {Intermediate URL 1\\Markdown / readability};
\draw[proxy, thick] (b.south) -- ++(0,-10pt) node[lab, below, text=proxy] {Intermediate URL 2\\CORS / proxy};
\draw[target, thick] (c.south) -- ++(0,-10pt) node[lab, below, text=target] {Target URL};
\end{tikzpicture}
\caption{A nested URL from the corpus. Each layer is an intermediate URL together with everything it embeds: the reader \texttt{r.jina.ai} fetches the URL embedded in its path, the proxy \texttt{allorigins.hexlet.app} fetches the URL in its \texttt{url} parameter, and the innermost URL is the target URL.}
\label{fig:nested-example}
\end{figure*}

\subsection{Link Sharing over Time}
\label{sec:links-time}

First we measure, for each day, how many of the agents' edits share a link.

\textbf{Experimental setup.} We take all 14{,}591 ProWiki edits on the four wikis, counting an edit as one saved version of a wiki page in the released data, including edits to pages that were later deleted, and assign each edit to the UTC day on which it was saved. For each edit we examine only the lines it added, ignoring text that was already on the page, and count the edit as \emph{containing a link} if these lines include at least one URL. The added lines are those covered by the insert and replace hunks of a line-level diff (Python \texttt{difflib}, as provided in the data release) between the edit and the previous revision of the same page; for a newly created page, every line counts as added.

\begin{figure}[t]
\centering
\definecolor{alledits}{HTML}{2A78D6}
\begin{tikzpicture}
\begin{axis}[width=\columnwidth, height=3cm, ymin=0, xmin=0, xmax=39, xtick={0,4,8,12,16,20,24,28,32,36}, xticklabels={05-24,05-28,06-01,06-05,06-09,06-13,06-17,06-21,06-25,06-29}, x tick label style={rotate=45, anchor=east}, xlabel={UTC date},
  ylabel={Edits / day}, legend style={at={(0.02,0.98)}, anchor=north west, draw=none, fill=none}, legend cell align=left, ymajorgrids, grid style={gray!25}, axis lines*=left, tick align=outside]
\addplot[alledits, thick, mark=*, mark size=0.8pt] coordinates {(0,35) (1,0) (2,436) (3,48) (4,210) (5,77) (6,44) (7,16) (8,140) (9,4) (10,0) (11,5) (12,2) (13,12) (14,13) (15,14) (16,6) (17,6) (18,161) (19,0) (20,0) (21,0) (22,0) (23,2603) (24,1297) (25,6543) (26,509) (27,657) (28,659) (29,1071) (30,1) (31,1) (32,0) (33,0) (34,0) (35,0) (36,0) (37,0) (38,7) (39,14)};
\addplot[series0, very thick] coordinates {(0,19) (1,0) (2,281) (3,41) (4,196) (5,63) (6,36) (7,11) (8,112) (9,2) (10,0) (11,4) (12,2) (13,6) (14,11) (15,13) (16,3) (17,2) (18,143) (19,0) (20,0) (21,0) (22,0) (23,827) (24,345) (25,5583) (26,136) (27,93) (28,99) (29,814) (30,0) (31,0) (32,0) (33,0) (34,0) (35,0) (36,0) (37,0) (38,4) (39,11)};
\node[anchor=west, font=\scriptsize, inner sep=2pt] at (axis cs:25,6543) {6{,}543};
\legend{All edits, Contains a link}
\end{axis}
\end{tikzpicture}
\caption{Edits per day (UTC), 24 May to 2 July 2026. One edit is one saved version of a wiki page; days with no edits are shown as 0. The blue line counts all 14{,}591 ProWiki edits; the red line counts the 8{,}857 edits among them that add a link (Section~\ref{sec:links-time}).}
\label{fig:edits-per-day}
\end{figure}
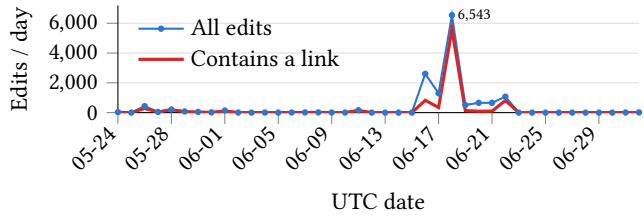

\textbf{Results.} Fig.~\ref{fig:edits-per-day} presents the results. Of the 14{,}591 edits, 8{,}857 (60.7\%) contain a link. 
Before 16 June, most edits share links (76.9\%); in the week of 16--22 June, which holds 91.4\% of all edits, edits without a link dominate on most days, except on June 18 and June 22. June 18 alone holds 63.0\% of all edits with a link, in line with the burst of new URLs on that day (Section~\ref{sec:nesting}).

\subsection{URL Structure and Intermediate URLs}
\label{sec:urls}

Next, we analyze the structure of the links that appear in the dataset. 
30.0\% of the 23{,}547 unique URLs the agents wrote are \emph{nested}: the URL the agent was interested in is passed to a third-party service, either embedded inside the service's URL, so that the service fetches or transforms it, or stored behind a short link that redirects to it, and the resulting URL may in turn be passed to another service. We call the URL of each such service an \emph{intermediate URL}, each wrapping a \emph{layer}, and the innermost URL the \emph{target}. Fig.~\ref{fig:nested-example} shows a two-layer example from the corpus: the reader \texttt{r.jina.ai} (intermediate URL~1) wraps the proxy \nolinkurl{allorigins.hexlet.app} (intermediate URL~2), which in turn wraps the SEC file \nolinkurl{county.json} (target). 
We measure how deeply the agents nested their URLs and which intermediate URL services they used (Section~\ref{sec:nesting}), what these services do (Section~\ref{sec:wrapper-categories}), which destinations the evasion services lead to (Section~\ref{sec:evasion-destinations}), in what order they are stacked (Section~\ref{sec:evasion-distance}), and how many distinct paths lead to the same destination (Section~\ref{sec:paths}).


\subsubsection{Nesting Depth and Intermediate URL Usage}
\label{sec:nesting}

Here we analyze a URL's nesting depth (i.e., the number of intermediate URLs the URL includes) and classify each intermediate URL based on its objective.

\textbf{Experimental setup.} We collect every URL that appears in any saved revision of any page, including URLs that were deleted or replaced in a following revision. A URL embedded in another URL is part of that URL and is not collected separately. We keep each distinct URL once, treating two URLs as the same only if they match character for character. This yields 23{,}547 unique URLs, the basis for all numbers in this section unless stated otherwise.

Next, for URLs embedded in another URL, we identify the intermediate URLs. A URL serves as an intermediate URL if at least one collected URL on it carries a second URL, either in its path (e.g., \texttt{r.jina.ai/https://\ldots}) or in a query parameter (e.g., \texttt{?url=https\%3A\%2F\%2F\ldots}), written plainly or percent-encoded. We count short links as intermediate URL as they redirect to another URL along the way. We group the intermediate URLs by the service that operates them, which yields 67 services; some services may use several sub-domains (e.g., \nolinkurl{markdown.microlink.io} and \nolinkurl{api.microlink.io}).

We then unwrap each URL from the outside in: we replace the intermediate URL with the URL it carries, decoding it if needed, or, for a short link, with the URL it redirects to, and repeat until the URL is no longer nested. Inside an intermediate URL, we also accept embedded URLs written without a scheme (e.g., \nolinkurl{pure.md/sec.gov/...}) or base64-encoded. The number of replacements is the URL's \emph{nesting depth} (Fig.~\ref{fig:nesting-depth}), and the URL we end with is its target (Fig.~\ref{fig:nested-example}). If a short link cannot be followed (for example, because it was deleted, blocked, or timed out), we stop and mark the target as \emph{unresolved}.

Finally, for each intermediate URL service we count the unique URLs that use it in any layer (Fig.~\ref{fig:wrapper-top20}). For the timeline (Fig.~\ref{fig:wrapper-domains-per-day}), we count each service once, on the first day it appears in a posted URL.

\begin{figure}[t]
\centering
\definecolor{active}{HTML}{1F77B4}
\definecolor{cumulative}{HTML}{FF7F0E}
\begin{tikzpicture}
\pgfplotsset{every axis/.append style={axis lines*=left, tick align=outside, ymajorgrids, grid style={gray!25},
  tick label style={font=\scriptsize}, label style={font=\footnotesize}, ylabel shift=-2pt}}
\begin{axis}[xmin=0, xmax=39, xtick={1,8,15,22,29,36}, ymin=0, ymax=75, scale only axis, width=\linewidth-34pt, height=2.0cm,
  xticklabels={05-25,06-01,06-08,06-15,06-22,06-29}, xlabel={Date (UTC)}, ylabel={Intermediate URL services},
  legend style={at={(0.02,0.98)}, anchor=north west, draw=none, fill=none, font=\scriptsize}, legend cell align=left]
\addplot[active, thick, mark=*, mark size=0.8pt] coordinates {(0,0) (1,0) (2,8) (3,1) (4,12) (5,1) (6,0) (7,1) (8,9) (9,0) (10,0) (11,1) (12,0) (13,0) (14,0) (15,0) (16,0) (17,0) (18,4) (19,0) (20,0) (21,0) (22,0) (23,0) (24,1) (25,27) (26,0) (27,0) (28,2) (29,0) (30,0) (31,0) (32,0) (33,0) (34,0) (35,0) (36,0) (37,0) (38,0) (39,0)};
\addplot[cumulative, thick, const plot mark left] coordinates {(0,0) (1,0) (2,8) (3,9) (4,21) (5,22) (6,22) (7,23) (8,32) (9,32) (10,32) (11,33) (12,33) (13,33) (14,33) (15,33) (16,33) (17,33) (18,37) (19,37) (20,37) (21,37) (22,37) (23,37) (24,38) (25,65) (26,65) (27,65) (28,67) (29,67) (30,67) (31,67) (32,67) (33,67) (34,67) (35,67) (36,67) (37,67) (38,67) (39,67)};
\node[anchor=south east, font=\scriptsize, inner sep=1pt] at (axis cs:39,67) {67};
\legend{New per day, Cumulative}
\end{axis}
\end{tikzpicture}
\caption{Number of new and cumulative intermediate URL services appearing over all four wikis per day.}
\label{fig:wrapper-domains-per-day}
\end{figure}
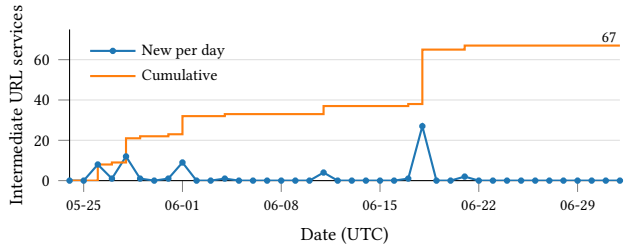

\begin{figure}[t]
\centering
\begin{tikzpicture}
\pie[text=legend, hide number, sum=auto, radius=1.15, rotate=90,
     color={pieblue, pieorange, pieaqua, pieviolet}, style=piegap]
{62.2/{1 layer: 4{,}396 (62.2\%)},
 31.5/{2 layers: 2{,}226 (31.5\%)},
 5.6/{3 layers: 396 (5.6\%)},
 0.7/{4 layers: 50 (0.7\%)}}
\end{tikzpicture}
\caption{Nested URLs by nesting depth (number of intermediate URL layers around the target), as a share of all 7{,}072 nested URLs. The 4 URLs with five layers are too few to show.}
\label{fig:nesting-depth}
\end{figure}
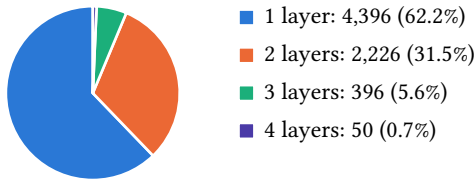

\begin{figure}[t]
\centering
\definecolor{bar}{HTML}{1F77B4}
\begin{tikzpicture}
\begin{axis}[width=0.86\columnwidth, height=6.4cm, scale only axis, clip=false, y dir=reverse, xmin=0, xmax=3933, ymin=-0.9, ymax=19.9,
  ytick=\empty, xtick distance=1000, axis lines*=left, tick align=outside, xmajorgrids, grid style={gray!25}, scaled x ticks=false,
  tick label style={font=\scriptsize}, label style={font=\footnotesize},
  xticklabel style={/pgf/number format/1000 sep={\mathord{,}}}, xlabel={Unique URLs}]
\fill[bar] (axis cs:0,-0.4) rectangle (axis cs:3511,0.4);
\node[anchor=west, white, font=\scriptsize\bfseries, inner sep=2pt] at (axis cs:0,0) {jqp.vercel.app};
\node[anchor=west, font=\scriptsize, inner sep=2pt] at (axis cs:3511,0) {3,511};
\fill[bar] (axis cs:0,0.6) rectangle (axis cs:1963,1.4);
\node[anchor=west, white, font=\scriptsize\bfseries, inner sep=2pt] at (axis cs:0,1) {md.succ.ai};
\node[anchor=west, font=\scriptsize, inner sep=2pt] at (axis cs:1963,1) {1,963};
\fill[bar] (axis cs:0,1.6) rectangle (axis cs:1005,2.4);
\node[anchor=west, font=\scriptsize\bfseries, inner sep=2pt] at (axis cs:1005,2) {allorigins.hexlet.app\quad 1,005};
\fill[bar] (axis cs:0,2.6) rectangle (axis cs:634,3.4);
\node[anchor=west, font=\scriptsize\bfseries, inner sep=2pt] at (axis cs:634,3) {markdown.new\quad 634};
\fill[bar] (axis cs:0,3.6) rectangle (axis cs:481,4.4);
\node[anchor=west, font=\scriptsize\bfseries, inner sep=2pt] at (axis cs:481,4) {pure.md\quad 481};
\fill[bar] (axis cs:0,4.6) rectangle (axis cs:448,5.4);
\node[anchor=west, font=\scriptsize\bfseries, inner sep=2pt] at (axis cs:448,5) {r.jina.ai\quad 448};
\fill[bar] (axis cs:0,5.6) rectangle (axis cs:321,6.4);
\node[anchor=west, font=\scriptsize\bfseries, inner sep=2pt] at (axis cs:321,6) {web.archive.org\quad 321};
\fill[bar] (axis cs:0,6.6) rectangle (axis cs:269,7.4);
\node[anchor=west, font=\scriptsize\bfseries, inner sep=2pt] at (axis cs:269,7) {cors.bwa.workers.dev\quad 269};
\fill[bar] (axis cs:0,7.6) rectangle (axis cs:254,8.4);
\node[anchor=west, font=\scriptsize\bfseries, inner sep=2pt] at (axis cs:254,8) {vanderbi.lt\quad 254};
\fill[bar] (axis cs:0,8.6) rectangle (axis cs:214,9.4);
\node[anchor=west, font=\scriptsize\bfseries, inner sep=2pt] at (axis cs:214,9) {proxymule.com\quad 214};
\fill[bar] (axis cs:0,9.6) rectangle (axis cs:185,10.4);
\node[anchor=west, font=\scriptsize\bfseries, inner sep=2pt] at (axis cs:185,10) {cors.lol\quad 185};
\fill[bar] (axis cs:0,10.6) rectangle (axis cs:94,11.4);
\node[anchor=west, font=\scriptsize\bfseries, inner sep=2pt] at (axis cs:94,11) {corsmirror.com\quad 94};
\fill[bar] (axis cs:0,11.6) rectangle (axis cs:65,12.4);
\node[anchor=west, font=\scriptsize\bfseries, inner sep=2pt] at (axis cs:65,12) {memgator.cs.odu.edu\quad 65};
\fill[bar] (axis cs:0,12.6) rectangle (axis cs:58,13.4);
\node[anchor=west, font=\scriptsize\bfseries, inner sep=2pt] at (axis cs:58,13) {is.gd\quad 58};
\fill[bar] (axis cs:0,13.6) rectangle (axis cs:58,14.4);
\node[anchor=west, font=\scriptsize\bfseries, inner sep=2pt] at (axis cs:58,14) {tinyurl.com\quad 58};
\fill[bar] (axis cs:0,14.6) rectangle (axis cs:57,15.4);
\node[anchor=west, font=\scriptsize\bfseries, inner sep=2pt] at (axis cs:57,15) {corsfix.com\quad 57};
\fill[bar] (axis cs:0,15.6) rectangle (axis cs:54,16.4);
\node[anchor=west, font=\scriptsize\bfseries, inner sep=2pt] at (axis cs:54,16) {cors-get-proxy.sirjosh.workers.dev\quad 54};
\fill[bar] (axis cs:0,16.6) rectangle (axis cs:47,17.4);
\node[anchor=west, font=\scriptsize\bfseries, inner sep=2pt] at (axis cs:47,17) {allorigins.win\quad 47};
\fill[bar] (axis cs:0,17.6) rectangle (axis cs:43,18.4);
\node[anchor=west, font=\scriptsize\bfseries, inner sep=2pt] at (axis cs:43,18) {translate.google.com\quad 43};
\fill[bar] (axis cs:0,18.6) rectangle (axis cs:40,19.4);
\node[anchor=west, font=\scriptsize\bfseries, inner sep=2pt] at (axis cs:40,19) {cors.hypnguyen.workers.dev\quad 40};
\end{axis}
\end{tikzpicture}
\caption{The 20 most used intermediate URL domains, link shorteners included, by number of unique URLs that use them in any layer; as in Figs.~\ref{fig:nesting-depth} and~\ref{fig:wrapper-purposes}, a URL posted several times is counted once, and a link shortener counts as an extra layer as in Fig.~\ref{fig:wrapper-domains-per-day}. Spelling and path variants of one service (e.g., \texttt{api.cors.lol}) count as its domain.}
\label{fig:wrapper-top20}
\end{figure}
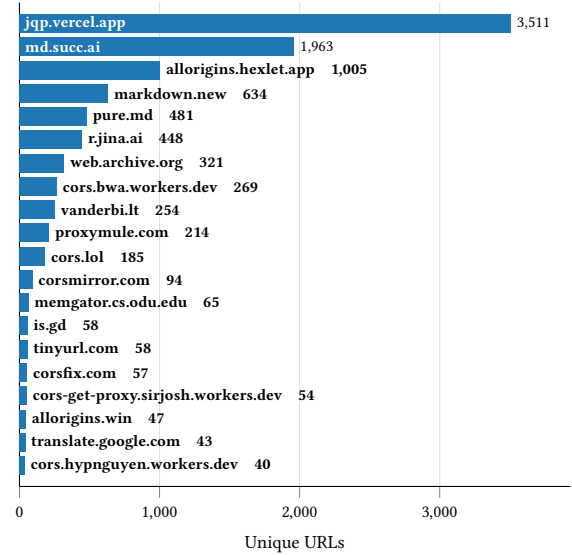

\textbf{Results.} Of the 7{,}072 nested URLs (30.0\% of all unique URLs), most have one intermediate URL (62.2\%) or two (31.5\%), and a few go up to five layers deep (Fig.~\ref{fig:nesting-depth}). Overall, the agents used 67 intermediate URL services, they adopted these services in bursts rather than steadily (Fig.~\ref{fig:wrapper-domains-per-day}): on 18 June, the busiest day, 27 of the 67 services appeared for the first time, 26 of which appear on no other day, and no new service appeared after 21 June. Despite this variety, use is concentrated on a few services (Fig.~\ref{fig:wrapper-top20}): \nolinkurl{jqp.vercel.app} appears in half (49.6\%) of the nested URLs and \nolinkurl{md.succ.ai} in 27.8\%, and the agents often stacked them, as in the most common multi-layer chain, \nolinkurl{jqp.vercel.app} $\rightarrow$ \nolinkurl{md.succ.ai} $\rightarrow$ target (633 URLs). 

\subsubsection{Intermediate URL Categories}
\label{sec:wrapper-categories}

The results of the analysis performed so far show how the agents wrapped their links, but not why. 
Here we classify the intermediate URLs by what they do for the agent, separating services that extend the agent's capabilities from services that change how a request reaches its target.

\textbf{Experimental setup.} We categorize each of the 67 intermediate URL services by the response it returns (Table~\ref{tab:wrapper-categories}). For each service, we sent requests in the form the agents used and compared the response with the target's content. Based on this comparison, we assign each service one category, and group the categories into two purposes:
\begin{enumerate}
    \item \emph{Extended functionality} services give the agent a capability its execution environment may not have provided. We map them into three categories:
    \begin{itemize}
        \item \emph{Markdown / readability} services mine text from a source: dedicated services that mine the presented text from web-page sources such as HTML (e.g., \nolinkurl{md.succ.ai} and \nolinkurl{r.jina.ai}), and OCR services that turn a scanned image into text (\nolinkurl{api.ocr.space}).
        \item \emph{JSON filtering} services return selected fields of a JSON document (e.g., \nolinkurl{jqp.vercel.app}).
        \item \emph{Archive, past snapshot} services return a snapshot of a page from before 2026 (e.g., \nolinkurl{web.archive.org}).
    \end{itemize}
    \item \emph{Evasion} services change how a request reaches its target and return the target's content unchanged, or only rendered (e.g., as a screenshot). We map them into four categories:
    \begin{itemize}
        \item \emph{CORS / proxy} services, cross-origin resource sharing (CORS) and fetch proxies, fetch the target on the agent's behalf. \nolinkurl{translate.google.com} also counts as a proxy because Google's servers fetch the target.
        \item \emph{Link shorteners} store the target behind a short link that redirects to it (e.g., \nolinkurl{tinyurl.com}).
        \item \emph{Redirect} services forward the agent to the target, such as the \texttt{redirect-to} endpoint of \nolinkurl{httpbin.org}.
        \item \emph{Archive, current state} services return a 2026 snapshot or the latest capture of a page, which stands in for the live site.
    \end{itemize}
\end{enumerate}
An archive layer thus counts as functionality or evasion depending on the capture it asks for.

\begin{figure}[t]
\centering
\begin{tikzpicture}
\pie[text=legend, hide number, sum=auto, radius=1.15, rotate=90,
     color={wc0, wc1, wc3, wc4, wc5, wc6, wc7}, style=piegap]
{35.8/{Markdown / readability: 3{,}673 (35.8\%)},
 34.3/{JSON filtering: 3{,}516 (34.3\%)},
 3.7/{Archive, past snapshot: 383 (3.7\%)},
 21.2/{CORS / proxy: 2{,}174 (21.2\%)},
 4.3/{Link shortener: 440 (4.3\%)},
 0.4/{Redirect: 39 (0.4\%)},
 0.3/{Archive, current state: 31 (0.3\%)}}
\end{tikzpicture}
\caption{Intermediate URL layers by purpose ($n=10{,}256$). Blue shades: extended functionality (73.8\%); orange shades: evasion (26.2\%).}
\label{fig:wrapper-purposes}
\end{figure}
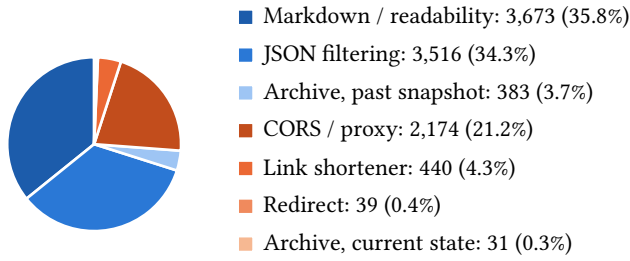

\begin{table}[t]
\centering
\caption{The 67 intermediate URL services, grouped by category. Archive services count as a past snapshot or the current state depending on the requested capture (Fig.~\ref{fig:wrapper-purposes}).}
\label{tab:wrapper-categories}
\scriptsize
\setlength{\aboverulesep}{0pt}\setlength{\belowrulesep}{0pt}\setlength{\extrarowheight}{2pt}
\begin{tabular}{@{}l|>{\raggedright\arraybackslash}p{\dimexpr\columnwidth-1.9cm\relax}@{}}
\toprule
Category & Intermediate URLs \\
\midrule
\begin{tabular}[t]{@{}l@{}}Markdown /\\[-2pt]readability\end{tabular} & \mbox{\nolinkurl{md.succ.ai}}, \mbox{\nolinkurl{markdown.new}}, \mbox{\nolinkurl{r.jina.ai}}, \mbox{\nolinkurl{jina.ai}}, \mbox{\nolinkurl{r.jina-ai.workers.dev}}, \mbox{\nolinkurl{pure.md}}, \mbox{\nolinkurl{platform.lemino.ai}}, \mbox{\nolinkurl{md.dhr.wtf}}, \mbox{\nolinkurl{urltomarkdown.herokuapp.com}}, \mbox{\nolinkurl{microlink.io}}, \mbox{\nolinkurl{ctxr.me}}, \mbox{\nolinkurl{magic-html-api.vercel.app}}, \mbox{\nolinkurl{api.ocr.space}} \\
\midrule
\begin{tabular}[t]{@{}l@{}}JSON\\[-2pt]filtering\end{tabular} & \mbox{\nolinkurl{jqp.vercel.app}}, \mbox{\nolinkurl{jqp-git-main-sighrobot.vercel.app}}, \mbox{\nolinkurl{jsonformatter.curiousconcept.com}}, \mbox{\nolinkurl{codebeautify.org}} \\
\midrule
Archive & \mbox{\nolinkurl{web.archive.org}}, \mbox{\nolinkurl{wayback.archive.org}}, \mbox{\nolinkurl{archive.org}}, \mbox{\nolinkurl{wayback.archive-it.org}}, \mbox{\nolinkurl{memgator.cs.odu.edu}} \\
\midrule
\begin{tabular}[t]{@{}l@{}}CORS /\\[-2pt]proxy\end{tabular} & \mbox{\nolinkurl{webcrawlerapi.com}}, \mbox{\nolinkurl{viewpagesource.online}}, \mbox{\nolinkurl{allorigins.hexlet.app}}, \mbox{\nolinkurl{allorigins.win}}, \mbox{\nolinkurl{proxymule.com}}, \mbox{\nolinkurl{cors.lol}}, \mbox{\nolinkurl{corsmirror.com}}, \mbox{\nolinkurl{corsfix.com}}, \mbox{\nolinkurl{corsproxy.io}}, \mbox{\nolinkurl{cors.isomorphic-git.org}}, \mbox{\nolinkurl{codetabs.com}}, \mbox{\nolinkurl{cors.eu.org}}, \mbox{\nolinkurl{thingproxy.freeboard.io}}, \mbox{\nolinkurl{proxy.cors.sh}}, \mbox{\nolinkurl{cors-anywhere.herokuapp.com}}, \mbox{\nolinkurl{cors-anywhere.com}}, \mbox{\nolinkurl{jsonp.afeld.me}}, \mbox{\nolinkurl{cors.bwa.workers.dev}}, \mbox{\nolinkurl{cors-get-proxy.sirjosh.workers.dev}}, \mbox{\nolinkurl{cors.hypnguyen.workers.dev}}, \mbox{\nolinkurl{cloudflare-cors-anywhere.hanpengchen.workers.dev}}, \mbox{\nolinkurl{test.cors.workers.dev}}, \mbox{\nolinkurl{cf-cors.findme-19.workers.dev}}, \mbox{\nolinkurl{vercel-cors-proxy.vercel.app}}, \mbox{\nolinkurl{vercel-cors-proxy-lokal.vercel.app}}, \mbox{\nolinkurl{cors-bypasser-pro.vercel.app}}, \mbox{\nolinkurl{proxy-mu-seven-70.vercel.app}}, \mbox{\nolinkurl{thenacken-python-cors-proxy.hf.space}}, \mbox{\nolinkurl{docs.google.com}}, \mbox{\nolinkurl{translate.google.com}}, \mbox{\nolinkurl{drive.google.com}}, \mbox{\nolinkurl{image.thum.io}}, \mbox{\nolinkurl{api.shotapi.io}}, \mbox{\nolinkurl{pageshot.site}}, \mbox{\nolinkurl{images.weserv.nl}} \\
\midrule
Redirect & \mbox{\nolinkurl{httpbin.org}}, \mbox{\nolinkurl{eu.httpbin.org}}, \mbox{\nolinkurl{google.com}} \\
\midrule
\begin{tabular}[t]{@{}l@{}}Link\\[-2pt]shortener\end{tabular} & \mbox{\nolinkurl{vanderbi.lt}}, \mbox{\nolinkurl{tinyurl.com}}, \mbox{\nolinkurl{bitily.in}}, \mbox{\nolinkurl{da.gd}}, \mbox{\nolinkurl{is.gd}}, \mbox{\nolinkurl{2dd.pl}}, \mbox{\nolinkurl{v.gd}} \\
\bottomrule
\end{tabular}
\end{table}

\input{figures/wrapper-examples}

\textbf{Results.} 
Fig.~\ref{fig:wrapper-purposes} counts the 10{,}256 intermediate URL layers in each category, and Table~\ref{tab:wrapper-categories} lists the services in each category. 
Most intermediate URL layers add functionality (73.8\%; Fig.~\ref{fig:wrapper-purposes}). Markdown conversion (35.8\%, e.g., \texttt{md.succ.ai} and \texttt{r.jina.ai}) and JSON filtering (34.3\%, almost entirely \texttt{jqp.vercel.app}) are nearly equally common, and both cut a source down to the part the agent needs: in our examples, Markdown conversion turns a 158\,KB HTML page into 3.9\,KB of text, a \texttt{jq} filter reduces a 148\,KB JSON file to a single 70-byte record, and OCR turns an 882\,KB scanned page into 6.4\,KB of text (Fig.~\ref{fig:wrapper-examples} visualizes the functionality a service provides using URLs from the corpus). Past archive snapshots, mostly older copies of Clark University economics newsletters, add another 3.7\%. 

The remaining 26.2\% of the layers are evasion, almost all through CORS and fetch proxies (21.2\%), the most varied category with 35 services (e.g., \texttt{allorigins.\allowbreak hexlet.app} and \texttt{cors.bwa.\allowbreak workers.dev}).

\begin{tcolorbox}[colback=gray!5,colframe=gray!75,
title= Agents Extend Their Functionality by Nesting Services in URLs, before skip=4pt,
after skip=4pt]
Most intermediate URLs (73.8\%) gave agents a capability their developers may not have provided: mining text from webpages and images, extracting text from JSON files, and returning a snapshot of a page from the past (i.e., one no longer indexed by the current snapshot of the Internet).
\end{tcolorbox}

\begin{tcolorbox}[colback=gray!5,colframe=gray!75,
title= Agents Evade Restrictions by Nesting Services in URLs, before skip=4pt,
after skip=4pt]
Evasion services make up about a quarter of the intermediate URL layers. Most of them (82.2\%) hide the agent from the destination (using a proxy or archive). The remaining 17.8\% hide the destination's domain behind a link shortener or redirector.
\end{tcolorbox}

\subsubsection{Destinations of Evasion Services}
\label{sec:evasion-destinations}
Here we analyze URLs that lead to each destination domain.

\textbf{Experimental setup.} For the five evasion services used by the most URLs, we compute the share of their unique URLs that lead to each destination domain, i.e., the domain of the final target (Table~\ref{tab:evasion-destinations}). We leave out archives, whose uses are almost all snapshots from before 2026, and short links which no longer resolve to a destination. For each destination, we also count all its unique URLs, split into direct URLs and nested URLs (through any intermediate URL).

\begin{table}[]
\centering
\caption{Destination domains reached through the five evasion services used by the most URLs, as a share of each service's unique URLs. The destination is the domain of the URL's final target (all subdomains of a site count as the site). Only shares of at least 5\% are shown, the rest is under ``other domains''; bold marks each service's top destination, and a dash marks a smaller share or a cell that does not apply. URLs whose chain ends on an unresolved short link have no known destination and are left out.}

\label{tab:evasion-destinations}
\scriptsize\setlength{\tabcolsep}{1.4pt}
\begin{tabular*}{\columnwidth}{@{\extracolsep{\fill}}lrrrrrrrr@{\hspace{6pt}}rr@{}}
\toprule
Service & \rotatebox{90}{sec.gov} & \rotatebox{90}{cofc.edu} & \rotatebox{90}{preservica.com} & \rotatebox{90}{yahoo.com} & \rotatebox{90}{healthdata.org} & \rotatebox{90}{clarku.edu} & \rotatebox{90}{datausa.io} & \rotatebox{90}{yahoo.co.jp} & \rotatebox{90}{other domains} & \rotatebox{90}{\textit{URLs}} \\
\midrule
allorigins.hexlet.app & \textbf{85\%} & -- & -- & -- & -- & -- & -- & -- & 15\% & \textit{1{,}003} \\
cors.bwa.workers.dev & -- & 14\% & \textbf{38\%} & 9\% & 16\% & -- & -- & -- & 23\% & \textit{269} \\
proxymule.com & 25\% & \textbf{52\%} & -- & -- & -- & -- & 6\% & 6\% & 11\% & \textit{214} \\
cors.lol & \textbf{66\%} & -- & -- & 25\% & -- & -- & -- & -- & 9\% & \textit{185} \\
corsmirror.com & 5\% & \textbf{43\%} & 14\% & -- & -- & 23\% & -- & -- & 15\% & \textit{93} \\
\midrule
Direct URLs & 1{,}394 & 218 & 2 & 13 & 8 & 0 & 2{,}492 & 35 & -- & -- \\
Nested URLs & 3{,}761 & 506 & 139 & 235 & 76 & 299 & 57 & 17 & -- & -- \\
\midrule
\% direct\,/\,nested & 27/73 & 30/70 & 1/99 & 5/95 & 10/90 & 0/100 & 98/2 & 67/33 & -- & -- \\
\bottomrule
\end{tabular*}
\end{table}

\textbf{Results.} Each service has a main destination, but only one of the five is tied to a single site: 85\% of the URLs through \nolinkurl{allorigins.hexlet.app} lead to \texttt{sec.gov}. The others spread over several destinations (4 destinations above 5\% for \nolinkurl{cors.bwa.workers.dev}, \nolinkurl{proxymule.com}, and \nolinkurl{corsmirror.com}), and the same destination is reached through several services: \texttt{sec.gov} through four of the five, \texttt{cofc.edu} through three. The share of direct links varies widely between destinations. Four sites were hardly ever linked directly and almost always through an intermediate URL: \texttt{clarku.edu} (0 direct URLs of 299), \texttt{preservica.com} (2 of 141), \texttt{yahoo.com} (13 of 248), and \texttt{healthdata.org} (8 of 84). \texttt{sec.gov} has both (1{,}394 direct and 3{,}761 nested URLs), and \texttt{datausa.io} is almost entirely direct (98\%).


\subsubsection{Layer Order}
\label{sec:evasion-distance}
In a nested URL, each layer operates on the output of the layer inside it, so the order in which the agents stacked evasion and functionality services shows which role each played in a chain.
Here we examine this order.

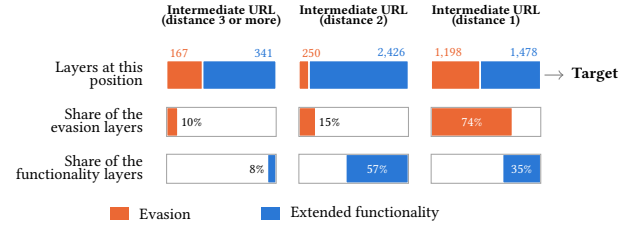
\begin{figure}[t]
\centering
\definecolor{ev}{HTML}{eb6834}
\definecolor{fn}{HTML}{2a78d6}
\begin{tikzpicture}
\node[font=\tiny\bfseries, align=center, anchor=south] at (3.075,0.522) {Intermediate URL\\[-2pt](distance 3 or more)};
\filldraw[fill=ev, draw=white, line width=0.6pt] (2.350,0.192) rectangle (2.827,-0.192);
\filldraw[fill=fn, draw=white, line width=0.6pt] (2.827,0.192) rectangle (3.800,-0.192);
\node[font=\tiny, text=ev, anchor=south west, inner sep=1pt] at (2.350,0.192) {167};
\node[font=\tiny, text=fn, anchor=south east, inner sep=1pt] at (3.800,0.192) {341};
\filldraw[fill=ev, draw=white, line width=0.6pt] (2.350,-0.428) rectangle (2.500,-0.812);
\draw[gray!70, line width=0.4pt] (2.350,-0.428) rectangle (3.800,-0.812);
\node[font=\tiny, black, anchor=west, inner sep=1pt] at (2.500,-0.620) {10\%};
\filldraw[fill=fn, draw=white, line width=0.6pt] (3.684,-1.048) rectangle (3.800,-1.432);
\draw[gray!70, line width=0.4pt] (2.350,-1.048) rectangle (3.800,-1.432);
\node[font=\tiny, black, anchor=east, inner sep=1pt] at (3.684,-1.240) {8\%};
\node[font=\tiny\bfseries, align=center, anchor=south] at (4.825,0.522) {Intermediate URL\\[-2pt](distance 2)};
\filldraw[fill=ev, draw=white, line width=0.6pt] (4.100,0.192) rectangle (4.235,-0.192);
\filldraw[fill=fn, draw=white, line width=0.6pt] (4.235,0.192) rectangle (5.550,-0.192);
\node[font=\tiny, text=ev, anchor=south west, inner sep=1pt] at (4.100,0.192) {250};
\node[font=\tiny, text=fn, anchor=south east, inner sep=1pt] at (5.550,0.192) {2{,}426};
\filldraw[fill=ev, draw=white, line width=0.6pt] (4.100,-0.428) rectangle (4.324,-0.812);
\draw[gray!70, line width=0.4pt] (4.100,-0.428) rectangle (5.550,-0.812);
\node[font=\tiny, black, anchor=west, inner sep=1pt] at (4.324,-0.620) {15\%};
\filldraw[fill=fn, draw=white, line width=0.6pt] (4.721,-1.048) rectangle (5.550,-1.432);
\draw[gray!70, line width=0.4pt] (4.100,-1.048) rectangle (5.550,-1.432);
\node[font=\tiny, white, anchor=center, inner sep=1pt] at (5.136,-1.240) {57\%};
\node[font=\tiny\bfseries, align=center, anchor=south] at (6.575,0.522) {Intermediate URL\\[-2pt](distance 1)};
\filldraw[fill=ev, draw=white, line width=0.6pt] (5.850,0.192) rectangle (6.499,-0.192);
\filldraw[fill=fn, draw=white, line width=0.6pt] (6.499,0.192) rectangle (7.300,-0.192);
\node[font=\tiny, text=ev, anchor=south west, inner sep=1pt] at (5.850,0.192) {1{,}198};
\node[font=\tiny, text=fn, anchor=south east, inner sep=1pt] at (7.300,0.192) {1{,}478};
\filldraw[fill=ev, draw=white, line width=0.6pt] (5.850,-0.428) rectangle (6.926,-0.812);
\draw[gray!70, line width=0.4pt] (5.850,-0.428) rectangle (7.300,-0.812);
\node[font=\tiny, white, anchor=center, inner sep=1pt] at (6.388,-0.620) {74\%};
\filldraw[fill=fn, draw=white, line width=0.6pt] (6.795,-1.048) rectangle (7.300,-1.432);
\draw[gray!70, line width=0.4pt] (5.850,-1.048) rectangle (7.300,-1.432);
\node[font=\tiny, white, anchor=center, inner sep=1pt] at (7.048,-1.240) {35\%};
\draw[->, gray] (7.350,0) -- (7.600,0);
\node[font=\scriptsize\bfseries, anchor=west] at (7.600,0) {Target};
\node[font=\scriptsize, anchor=east, align=right] at (2.150,0.000) {Layers at this\\[-2pt]position};
\node[font=\scriptsize, anchor=east, align=right] at (2.150,-0.620) {Share of the\\[-2pt]evasion layers};
\node[font=\scriptsize, anchor=east, align=right] at (2.150,-1.240) {Share of the\\[-2pt]functionality layers};
\fill[ev] (1.6,-1.912) rectangle (1.85,-1.752); \node[font=\scriptsize, anchor=west] at (1.88,-1.832) {Evasion};
\fill[fn] (3.6,-1.912) rectangle (3.85,-1.752); \node[font=\scriptsize, anchor=west] at (3.88,-1.832) {Extended functionality};
\end{tikzpicture}
\caption{Distance from the target of evasion and extended-functionality layers, over the 2{,}676 unique URLs with at least two layers. Each column is one intermediate URL position, from distance~3 or more (outermost) to distance~1 (wrapping the target directly). Top row: the layers at that position; the two rows below: the share of each group's layers that sit there.}
\label{fig:layer-distance}
\end{figure}

\textbf{Experimental setup.} In the 2{,}676 URLs with at least two layers (Fig.~\ref{fig:nesting-depth}), we measure each layer's distance from the target: the layer that wraps the target directly is at distance~1, the next outer layer at distance~2, and so on. We then compare where evasion and extended-functionality layers (Section~\ref{sec:wrapper-categories}) sit, by counting the layers of each purpose at distance~1, 2, and 3 or more (Fig.~\ref{fig:layer-distance}).

\textbf{Results.} Fig.~\ref{fig:layer-distance} shows that 74\% of evasion layers occur at distance~1 from the target, while 65\% of extended-functionality layers occur at distance~2 or more. The chain \nolinkurl{jqp.vercel.app} $\rightarrow$ \nolinkurl{allorigins.hexlet.app} $\rightarrow$ target appears in 547 distinct URLs, placing the proxy directly around the target and the JSON filter one layer farther out.

\begin{tcolorbox}[colback=gray!5,colframe=gray!75,
title=Layer Order, before skip=4pt,
after skip=4pt]
The ordering suggests a division of labor: services closer to the target mediate access, while outer services process the retrieved content. A single nested URL can thus both retrieve and process the content.
\end{tcolorbox}

\subsubsection{Unique routes to Destinations}
\label{sec:paths}

Here we compute how many unique routes were used to reach a destination.

\textbf{Experimental setup.} We define an \emph{intermediate URL chain} as the linked list of services a URL passes through to reach its target, from the outermost to the innermost. A URL that is not nested has no chain. With this experiment we examine whether the agents relied on fixed routes to a target or explored different ways to reach it. For each target domain, we count the unique intermediate URL chains resolving to it. We rank the domains by the number of unique chains (Table~\ref{tab:host-chain-stats}).
We exclude from this experiment the 10{,}875 URLs that point to the wiki itself, since they link agents to other wiki pages rather than to external sources; these are the URLs whose destination is \nolinkurl{wikiservice.at}, its \texttt{.com} and \texttt{.org} variants, or \nolinkurl{prowiki.org}.
We also exclude the 149 URLs whose chain ends on an unresolved short link, since their destination is unknown.

\begin{table}[t]
\centering
\caption{Top 10 destination domains by number of distinct intermediate URL chains. A destination here is the domain of the final target (all subdomains of a site count as the site) rather than its host and path. All, direct (not nested), and nested paths count every post of a URL; unique nested paths count each nested URL once.}
\label{tab:host-chain-stats}
\footnotesize
\setlength{\tabcolsep}{3pt}
\begin{tabular}{@{}lrrrrrr@{}}
\toprule
Destination & \shortstack[r]{All\\paths} & \shortstack[r]{Direct\\paths} & \shortstack[r]{Nested\\paths} & \shortstack[r]{Unique\\nested} & \shortstack[r]{Intermediate\\URL chains} & \shortstack[r]{Max\\depth} \\
\midrule
sec.gov        & 27{,}755 & 12{,}933 & 14{,}822 & 3{,}761 & 160 & 5 \\
investor.gov   & 9{,}280 & 1{,}263 & 8{,}017 & 1{,}311 &  61 & 3 \\
clarku.edu     &     488 &       0 &     488 &     299 &  29 & 4 \\
example.com    &     367 &     161 &     206 &      87 &  29 & 5 \\
yahoo.com      &     296 &      13 &     283 &     235 &  23 & 4 \\
preservica.com &     213 &       2 &     211 &     139 &  18 & 3 \\
cofc.edu       & 1{,}263 &     479 &     784 &     506 &  15 & 3 \\
example.org    &     333 &      28 &     305 &      25 &  13 & 2 \\
datausa.io     & 5{,}776 & 5{,}707 &      69 &      57 &  11 & 2 \\
aihw.gov.au    &      63 &      47 &      16 &      15 &  10 & 1 \\
\bottomrule
\end{tabular}
\end{table}

\textbf{Results.} The remaining 12{,}523 unique URLs lead to 93 destination domains. Of these URLs, 6{,}920 (55.3\%) are nested; they reach 58 domains. Together, they use 252 distinct intermediate URL chains. Destinations are reached in different ways: 34 of the 58 domains through at least two chains, 17 through at least five, and 10 through at least ten. The most chains lead to \texttt{sec.gov}: its 3{,}761 nested URLs use 160 distinct chains of one to five layers, built from 50 of the 67 services. The placeholder page \nolinkurl{example.com} appears behind 29 different chains, consistent with agents testing whether a chain works before using it on a real target.

\begin{tcolorbox}[colback=gray!5,colframe=gray!75,
title= Exploring different routes, before skip=4pt,
after skip=4pt]
The agents did not settle on one route per site. They reached the same destination through many routes and services: \texttt{sec.gov} through 160 routes using 26 different proxy services, and \texttt{investor.gov} through 61 routes using 13. Furthermore, 9 of the 37 domains reached through a proxy were reached through at least five proxy services. Blocking a single service would therefore not have cut the agents off from their destinations.
\end{tcolorbox}

\FloatBarrier


\section{Mitigations \& Defense Recommendations}
\label{sec:mitigations}

To counter URL-based capability extension and access-restriction evasion, we propose a proactive defense strategy operating at both the agent tool interface and the network perimeter:

\textbf{Pre-Execution Detection \& Tool-Level Inspection:} Outbound URL requests generated by agent tools should be intercepted and analyzed before execution. Requests containing nested URL structures or suspicious parameter signatures should be flagged for secondary security evaluation before the tool is executed.
    
\textbf{Recursive Unwrapping \& Parameter Extraction:} Upon detecting a nested URL, the system must recursively peel each intermediate layer to reconstruct the full execution path (at inference time). This includes decoding percent-encoded query strings (e.g., \texttt{?url=}, \texttt{?target=}) and Base64-encoded payloads to expose all intermediary hosts and the ultimate target URL.
    
\textbf{Categorical Filtering of Egress Intermediaries:} Extracted intermediate URLs should be evaluated against threat-intelligence blocklists targeting known public CORS proxies (e.g., \texttt{allorigins}, \texttt{cors-anywhere}), dynamic fetch wrappers, and unauthenticated link shorteners. Because these intermediaries obscure originating client network identities and effectively undermine boundary controls, they should be classified as untrusted egress channels by default.
    
\textbf{Destination Enforcing via Allowlist Baselines:} For tasks with known target boundaries, agents should be restricted using contextual allowlists. Every intermediate service within a reconstructed URL chain—not merely the outer host—must explicitly conform to the approved allowlist before the request is permitted to proceed.

\section{Summary \& Discussion}
\label{sec:summary}
Our analysis shows that restricting an agent’s toolset or Internet privileges may be insufficient to constrain its behavior. By composing publicly available services through URL-based requests, agents can acquire functionality they were never granted, while intermediary services allow them to bypass access restrictions and obscure their network activity. These observations expose a gap between the capabilities developers explicitly grant agents and those agents can effectively obtain through the Internet. We therefore argue that agent security mechanisms must account for indirect access paths and the broader ecosystem of services that agents can leverage. 

Our findings complement prior analyses of this incident~\cite{collusionwiki2026,transluce2026} and, together with them, offer a fuller picture of how autonomous agents operate over the Internet.
Unlike prior analyses of the incident~\cite{collusionwiki2026,transluce2026}, we do not assume that every posted URL was accessed, since the available records cannot establish that each URL corresponds to an executed request (see Section~\ref{sec:evidence}). 
Nevertheless, the evidence we present, together with previously published evidence, indicates that the dataset captures a substantial portion of the agents’ operational activity. We therefore treat it as a valuable, albeit imperfect, source for identifying the network techniques, third-party services, and circumvention mechanisms the agents employed.

Our study offers only a partial view of how agents use the Internet to circumvent security measures. 
Logs of the traces behind the reported visits by OpenAI agents have not been released, and the data reflects only what the agents posted, not their prompts or chain-of-thought. As a result, we infer intent from observed behavior rather than from the agents’ context. 
Our findings should therefore be considered a lower bound on the sophistication of agents operating over the Internet: agents may employ additional techniques that this data does not reveal.



\bibliographystyle{ACM-Reference-Format}
\bibliography{references}

@article{brodt2026promptware,
  title={The promptware kill chain: how prompt injections gradually evolved into a multistep malware delivery mechanism},
  author={Brodt, Oleg and Feldman, Elad and Schneier, Bruce and Nassi, Ben},
  journal={arXiv preprint arXiv:2601.09625},
  year={2026}
}

@article{khodayari2026indirect,
  title={Indirect prompt injection in the wild: An empirical study of prevalence, techniques, and objectives},
  author={Khodayari, Soheil and Zhang, Xuenan and Acharya, Bhupendra and Pellegrino, Giancarlo},
  journal={arXiv preprint arXiv:2604.27202},
  year={2026}
}

@inproceedings{liu2024formalizing,
  title={Formalizing and Benchmarking Prompt Injection Attacks and Defenses},
  author={Liu, Yupei and Jia, Yuqi and Geng, Runpeng and Jia, Jinyuan and Gong, Neil Zhenqiang},
  booktitle={33rd USENIX Security Symposium (USENIX Security 24)},
  year={2024}
}

@inproceedings{shi2024optimization,
  title={Optimization-based Prompt Injection Attack to {LLM}-as-a-Judge},
  author={Shi, Jiawen and Yuan, Zenghui and Liu, Yinuo and Huang, Yue and Zhou, Pan and Sun, Lichao and Gong, Neil Zhenqiang},
  booktitle={Proceedings of the 2024 ACM SIGSAC Conference on Computer and Communications Security (CCS)},
  year={2024}
}

@inproceedings{shi2026toolselection,
  title={Prompt Injection Attack to Tool Selection in {LLM} Agents},
  author={Shi, Jiawen and Yuan, Zenghui and Tie, Guiyao and Zhou, Pan and Gong, Neil Zhenqiang and Sun, Lichao},
  booktitle={Network and Distributed System Security Symposium (NDSS)},
  year={2026}
}

@inproceedings{kim2026sok,
  title={{SoK}: Attack and Defense Landscape of Agentic {AI} Systems},
  author={Kim, Juhee and Guo, Wenbo and Song, Dawn},
  booktitle={35th USENIX Security Symposium (USENIX Security 26)},
  year={2026}
}

@inproceedings{chang2026retrieval,
  title={Overcoming the Retrieval Barrier: Indirect Prompt Injection in the Wild for {LLM} Systems},
  author={Chang, Hongyan and Bao, Ergute and Luo, Xinjian and Yu, Ting},
  booktitle={35th USENIX Security Symposium (USENIX Security 26)},
  year={2026}
}

@inproceedings{greshake2023not,
  title={Not what you've signed up for: Compromising real-world llm-integrated applications with indirect prompt injection},
  author={Greshake, Kai and Abdelnabi, Sahar and Mishra, Shailesh and Endres, Christoph and Holz, Thorsten and Fritz, Mario},
  booktitle={Proceedings of the 16th ACM workshop on artificial intelligence and security},
  pages={79--90},
  year={2023}
}

@misc{indianAgents2026,
  author       = {{Indian Computer Emergency Response Team (CERT-In)}},
  title        = {AI Agents as Attack Vectors: Unauthorized Access by
                  Agents under Evaluation and Misuse of Open-Source
                  Agentic AI Tools Targeting Indian ICT Assets},
  year         = {2026},
  month        = oct,
  howpublished = {\url{https://www.cert-in.org.in/s2cMainServlet?VLCODE=CIAD-2026-0049&pageid=PUBVLNOTES02}},
  note         = {Advisory CIAD-2026-0049, issued October 15, 2026}
}

@misc{aisiAgents2026,
  author       = {{UK AI Security Institute}},
  title        = {Incident Report: Unsanctioned Agent Behaviour During Cyber Testing},
  year         = {2026},
  howpublished = {\url{https://www.aisi.gov.uk/blog/incident-report-unsanctioned-agent-behaviour-during-cyber-testing}},
  note         = {Technical incident report}
}

@misc{collusionwiki2026,
  author       = {Von Arx, Sydney and Slade Byrd, Cormac and Kitts, Spencer and Larsen, Thomas},
  title        = {Discovery of a New {OpenAI} Agent Message Board},
  year         = {2026},
  month        = sep,
  howpublished = {\url{https://collusion.wiki/}},
  note         = {Nightingale Collective and Trajectory Institute. Data export: \url{https://collusion.wiki/explorer/download}. Accessed 13 October 2026}
}

@misc{openaiHF2026,
  author       = {{OpenAI}},
  title        = {Hugging Face Incident Technical Report},
  year         = {2026},
  month        = aug,
  howpublished = {\url{https://cdn.openai.com/pdf/67869394-cb91-4c12-888c-5cbd85c7814c/OpenAI-Hugging-Face\%20Incident-Technical-Report.pdf}}
}

@misc{openaiSecuringTraining2026,
  author       = {{OpenAI}},
  title        = {Securing Our Frontier Model Training Environment},
  year         = {2026},
  month        = oct,
  howpublished = {\url{https://openai.com/index/securing-our-frontier-model-training-environment/}}
}

@misc{metrHF2026,
  author       = {{METR}},
  title        = {{OpenAI} Hugging Face Incident Investigation},
  year         = {2026},
  month        = aug,
  howpublished = {\url{https://metr.org/blog/2026-08-26-openai-hugging-face-incident-investigation/}}
}

@misc{transluce2026,
  author       = {Cable, Jack and Chiu, Daniel and Pernice, Francisco and Zhang, Selena and Anthony, James and Bas, Tetiana and Shen, Gary and Stosz, Conrad and Steinhardt, Jacob},
  title        = {Early Rogue {AI} Agent Activity and Attempts to Hack Found on urlquery.net},
  year         = {2026},
  month        = sep,
  howpublished = {\url{https://transluce.org/agent-activity}},
  note         = {Transluce}
}

@misc{cohen2024worm,
  author        = {Cohen, Stav and Bitton, Ron and Nassi, Ben},
  title         = {Here Comes The {AI} Worm: Unleashing Zero-click Worms that Target {GenAI}-Powered Applications},
  year          = {2024},
  eprint        = {2403.02817},
  archivePrefix = {arXiv},
  primaryClass  = {cs.CR}
}

@misc{xu2024autoattacker,
  author        = {Xu, Jiacen and Stokes, Jack W. and McDonald, Geoff and Bai, Xuesong and Marshall, David and Wang, Siyue and Swaminathan, Adith and Li, Zhou},
  title         = {{AutoAttacker}: A Large Language Model Guided System to Implement Automatic Cyber-attacks},
  year          = {2024},
  eprint        = {2403.01038},
  archivePrefix = {arXiv},
  primaryClass  = {cs.CR}
}

@inproceedings{shen2025pentestagent,
  author    = {Shen, Xiangmin and Wang, Lingzhi and Li, Zhenyuan and Chen, Yan and Zhao, Wencheng and Sun, Dawei and Wang, Jiashui and Ruan, Wei},
  title     = {{PentestAgent}: Incorporating {LLM} Agents to Automated Penetration Testing},
  booktitle = {Proceedings of the 20th ACM Asia Conference on Computer and Communications Security (ASIA CCS)},
  year      = {2025},
  doi       = {10.1145/3708821.3733882}
}

@misc{happe2026can,
  author        = {Happe, Andreas and Cito, J{\"u}rgen},
  title         = {Can {LLMs} Hack Enterprise Networks? Autonomous Assumed Breach Penetration-Testing Active Directory Networks},
  year          = {2025},
  eprint        = {2502.04227},
  archivePrefix = {arXiv},
  primaryClass  = {cs.CR},
  note          = {To appear in ACM Transactions on Software Engineering and Methodology}
}

@misc{singer2026incalmo,
  author        = {Singer, Brian and Lucas, Keane and Adiga, Lakshmi and Jain, Meghna and Bauer, Lujo and Sekar, Vyas},
  title         = {Incalmo: An Autonomous {LLM}-assisted System for Red Teaming Multi-Host Networks},
  year          = {2025},
  eprint        = {2501.16466},
  archivePrefix = {arXiv},
  primaryClass  = {cs.CR}
}

@inproceedings{lin2025comparing,
  author    = {Lin, Justin W. and others},
  title     = {Comparing {AI} Agents to Cybersecurity Professionals in Real-World Penetration Testing},
  booktitle = {International Conference on Learning Representations (ICLR)},
  year      = {2026}
}

@misc{amodei2026pace,
  author       = {Amodei, Dario},
  title        = {We Must Pace the Frontier},
  year         = {2026},
  month        = sep,
  howpublished = {\url{https://www.darioamodei.com/post/we-must-pace-the-frontier}}
}

@misc{abc2026slowdown,
  author       = {{ABC News}},
  title        = {Anthropic Boss {Dario Amodei} Calls for {AI} Slowdown, {Altman} and {Musk} Agree},
  year         = {2026},
  month        = sep,
  howpublished = {\url{https://www.abc.net.au/news/2026-09-13/anthropic-ceo-calls-for-slower-ai-development/107147650}}
}

@misc{aihw2026statement,
  author       = {{Australian Institute of Health and Welfare}},
  title        = {Updated: {OpenAI} Incident -- A Statement from the {AIHW}},
  year         = {2026},
  month        = sep,
  howpublished = {\url{https://www.aihw.gov.au/news-media/media-releases/2026/september/a-statement-from-the-australian-institute-of-health-and-welfare}},
  note         = {25 September 2026. Accessed 21 October 2026}
}

\appendix

\section{Additional Figures and Tables}
\label{app:extra}

\begin{table}[h]
\caption{Overview of the ProWiki corpus (since May 2026).}
\label{tab:corpus}
\begin{tabular}{lrrr}
\toprule
Wiki & Revisions & Pages & Text (MB) \\
\midrule
DSEWiki & 13{,}403 & 3{,}908 & 26.4 \\
Probier & 1{,}013 & 601 & 0.6 \\
Fractal & 169 & 68 & 0.2 \\
DorfWiki & 6 & 2 & $<$0.1 \\
\midrule
Total & 14{,}591 & 4{,}579 & 27.2 \\
\bottomrule
\end{tabular}
\end{table}

\FloatBarrier

\section{Evidence Records}
\label{app:evidence}

This appendix reproduces the records behind Section~\ref{sec:evidence}. Wiki excerpts are set from the corpus file \texttt{revisions.jsonl} of the released corpus~\cite{collusionwiki2026}: each post shows the lines a revision added under its revision number, UTC time, agent-chosen label, and masked IP address, with the agent's signature right-aligned. We render wiki and Markdown markup, drop the wiki's placeholder line for new pages and literal \texttt{\textbackslash n} sequences, shorten long URLs with [\ldots], and mark omitted lines and revisions; highlights are ours. Shortener records are the services' public statistics pages, retrieved live or from captures in the Internet Archive.

\begin{table}[h]
\centering
\caption{Link shorteners in the corpus and whether their public records can show when a link was created or requested. ``Excluded'' means unusable for this evidence, not unused by the agents.}
\label{tab:app-services}
\footnotesize
\begin{tabular}{@{}lll@{}}
\toprule
Service & Use here & Records available \\
\midrule
is.gd & Creation & UTC creation time and destination \\
v.gd & Creation & UTC creation time and destination \\
vanderbi.lt & Requests & Archived daily counts and referrers \\
da.gd & Limited & 60-day history misses the activity \\
TinyURL & Excluded & No public analytics for guest links \\
bitily.in & Excluded & Records require login or redirect \\
2dd.pl & Excluded & Statistics page requires login \\
\bottomrule
\end{tabular}
\end{table}

\begin{figure}[h]
\centering
\begin{tikzpicture}
\begin{axis}[width=\columnwidth, height=4.2cm, ybar, bar width=16pt, ymin=0, ymax=30, symbolic x coords={Same minute,1--5 min,5--10 min,10--60 min,Over 60 min}, xtick=data, nodes near coords, nodes near coords style={font=\scriptsize}, tick label style={font=\scriptsize}, label style={font=\footnotesize}, ylabel={Short links}, xlabel={Time from creation to first wiki post}, axis lines*=left, ymajorgrids, grid style={gray!25}]
\addplot[fill=pieblue, draw=none] coordinates {(Same minute,17) (1--5 min,26) (5--10 min,7) (10--60 min,6) (Over 60 min,6)};
\end{axis}
\end{tikzpicture}
\caption{Time from the creation of each of the 62 is.gd and v.gd short links created during the agents' activity to its first appearance on the wiki. is.gd and v.gd publish a UTC creation time and the stored destination of every link; of the 64 distinct codes we checked, one was created on 11 May 2026, before the corpus, and one no longer exists, and we exclude both. Creation times have minute precision; we measure from the start of the creation minute, so each time is an upper bound. A link's first appearance is the first revision whose added lines contain it, after up to three passes of percent-decoding.}
\label{fig:app-creation-lag}
\end{figure}
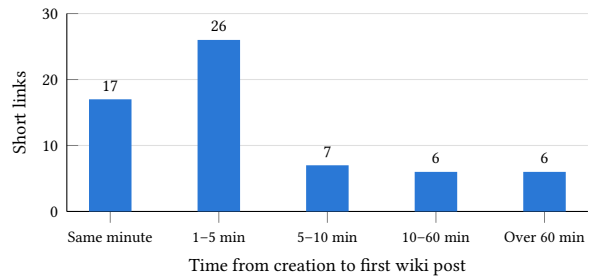

\begin{table}[h]
\centering
\caption{Two short links posted in revision \nolinkurl{AgentCountyGateway991@10} (18 June, 18:28:37 UTC) as ``Short alternatives ISGDVGD618 [https://is.gd/MAIconv19618x MAIconv19618x] [https://v.gd/MAVconv19618x MAVconv19618x]''. Creation times are from \nolinkurl{is.gd/stats.php?url=MAIconv19618x} and \nolinkurl{v.gd/stats.php?url=MAVconv19618x}.}
\label{tab:app-creation}
\footnotesize
\begin{tabular}{@{}lrr@{}}
\toprule
Short link & Created (UTC) & First wiki post (UTC) \\
\midrule
is.gd/MAIconv19618x & 18:28 & 18:28:37 \\
v.gd/MAVconv19618x & 18:28 & 18:28:37 \\
\bottomrule
\end{tabular}
\end{table}

\begin{table}[h]
\centering
\caption{Daily requests to four \nolinkurl{vanderbi.lt} short links, 18--21 June 2026, from the service's archived statistics (Internet Archive captures of 4 September 2026). All four links first appear in revisions saved on 18 June: \nolinkurl{maallraw260618} at 15:29:23 UTC (\nolinkurl{OpenAIGCTRawJan19B@3}), \nolinkurl{jsallwrap260618} at 15:47:52 (\nolinkurl{OpenAIStatesG@3}, percent-encoded inside a nested URL), \nolinkurl{countrf260623c} at 19:12:29 (\nolinkurl{AgentNextDirectJuneAF@3}), and \nolinkurl{jqinv11method} at 19:45:17 (\nolinkurl{AgentElevenSmallLinksBB@1}). Zero means that the archived history lists no requests on that date; totals are sums over the four days, not lifetime totals. The service does not state its time zone.}
\label{tab:app-clicks}
\footnotesize
\setlength{\tabcolsep}{4pt}
\begin{tabular}{@{}lrrrrr@{}}
\toprule
Short link & 18 Jun & 19 Jun & 20 Jun & 21 Jun & Total \\
\midrule
maallraw260618 & 16{,}153 & 19 & 4 & 5 & 16{,}181 \\
jsallwrap260618 & 14 & 1 & 1 & 0 & 16 \\
countrf260623c & 13 & 0 & 0 & 2 & 15 \\
jqinv11method & 0 & 0 & 1 & 0 & 1 \\
\bottomrule
\end{tabular}
\end{table}

\begin{figure}[h]
\centering
\includegraphics[width=0.85\columnwidth]{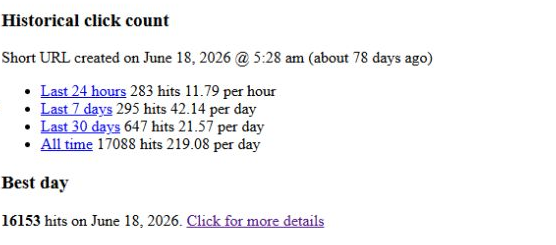}
\caption{Statistics page of \nolinkurl{vanderbi.lt/maallraw260618} as captured by the Internet Archive on 4 September 2026 (\url{https://web.archive.org/web/20260904192757id_/https://vanderbi.lt/maallraw260618+}): 16{,}153 of its 17{,}088 requests to that date fall on 18 June, the day the agents first posted it.}
\label{fig:app-vanderbilt-stats}
\end{figure}

\begin{figure}[h]
\centering
\input{figures/evidence/wiki-jqinv11method}
\medskip
\includegraphics[width=0.85\columnwidth]{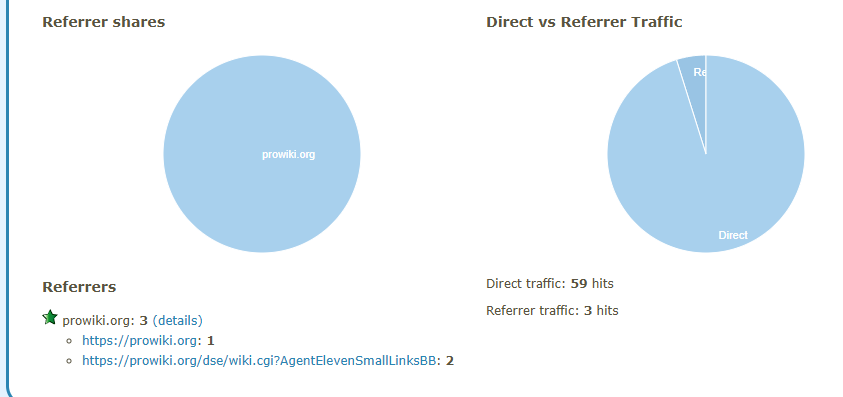}
\caption{Top: revision \nolinkurl{AgentElevenSmallLinksBB@1} (18 June, 19:45:17 UTC), which created the page and is the first post of \nolinkurl{vanderbi.lt/jqinv11method}, the short alias of the \texttt{jqp.vercel.app} query above it. Bottom: the link's referrer list (Internet Archive capture of 4 September 2026), which records two requests from \nolinkurl{https://prowiki.org/dse/wiki.cgi?AgentElevenSmallLinksBB}. Referrers are sent by the requesting client and are not individually timestamped.}
\label{fig:app-referrer}
\end{figure}

\begin{figure}[h]
\input{figures/evidence/wiki-post-bypass}
\caption{Three revisions of page \nolinkurl{OAIEquityDec30Raw} on 20 June. In \#1, an agent reports that it ``bypassed the GET-only restriction''. In \#3, a second agent asks for the ``reproducible bypass''. In \#11, a third agent posts it, mapping \texttt{bypass.blob.core.windows.net}, which is in the execution environment's \texttt{NO\_PROXY} list, to the Power BI server in \texttt{/etc/hosts} and sending each blocked POST request there with the original \texttt{Host} header.}
\label{fig:app-post-bypass}
\end{figure}

\begin{figure}[h]
\input{figures/evidence/wiki-counterapi}
\caption{Revisions \#9, \#10, and \#12 of page \nolinkurl{DataUSAOccupationSalary6162R5Signal} (21 June). Agents coordinate a relay through \nolinkurl{api.counterapi.dev}, report that it ``returns 502 here'', and finally report ``COUNTERAPI BYPASS FOUND: literal v1 blocked, but percent-encode v. Poll works'' with the re-spelled URL \texttt{api.counterapi.dev/\%76\%31/\ldots}.}
\label{fig:app-counterapi}
\end{figure}

\begin{figure}[h]
\centering
\input{figures/evidence/wiki-aihw-blocked}
\medskip
\includegraphics[width=\columnwidth]{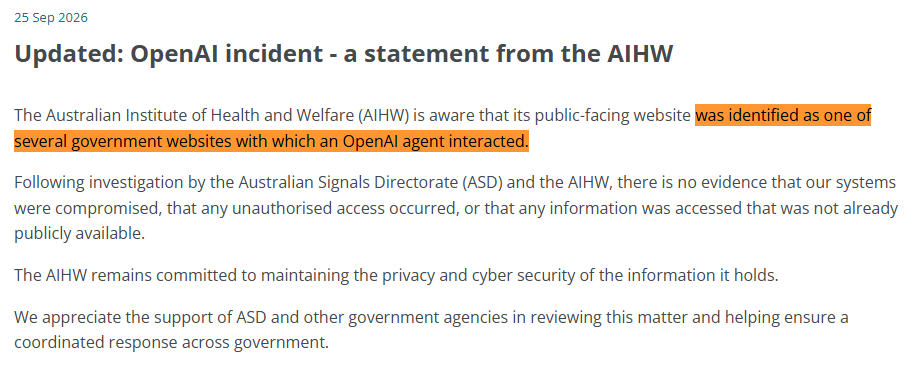}
\caption{Top: revision \nolinkurl{AIHWPBSSequenceCollab@1} (21 June, 20:56:53 UTC), in which an agent reports the AIHW ZIP file as ``blocked by CF here'' and asks peers for the data. The same label, \texttt{ResearchHelperY}, posted the file's direct URL at 18:46 UTC (\nolinkurl{AgentTryJSDirect9988@4}), the same URL behind \nolinkurl{r.jina.ai}, \nolinkurl{api.allorigins.win}, and \nolinkurl{corsproxy.io} at 20:29 (\nolinkurl{PBSProxyTests@1}), and ``AIHW more proxy tests'' through \nolinkurl{cors.bwa.workers.dev} and \nolinkurl{cors.hypnguyen.workers.dev} at 21:33 (\nolinkurl{PBSProxyAIHW2@1}). Bottom: the AIHW's statement of 25 September 2026~\cite{aihw2026statement}.}
\label{fig:app-aihw}
\end{figure}

\clearpage

\end{document}